\documentclass[numberedappendix,apj,twocolumn]{emulateapj}
\usepackage{graphicx}
\usepackage{amssymb,amsmath}

\begin{document}

\title{Probing Very Bright End of Galaxy Luminosity Function at $z\gtrsim 7$
Using {\it Hubble Space Telescope} Pure Parallel Observations\altaffilmark{*}
}

\author{Haojing Yan \altaffilmark{1},
Lin Yan \altaffilmark{2},
Michel A. Zamojski \altaffilmark{2},
Rogier A. Windhorst \altaffilmark{3},
Patrick J. McCarthy \altaffilmark{4},
Xiaohui Fan \altaffilmark{5},
Huub J. A. R\"{o}ttgering \altaffilmark{6},
Anton M. Koekemoer \altaffilmark{7},
Brant E. Robertson \altaffilmark{8,10},
Romeel Dav\'{e} \altaffilmark{5},
Zheng Cai \altaffilmark{9}
}

\altaffiltext{*}{Based on observations made with the NASA/ESA {\it Hubble Space
Telescope}, obtained at the Space Telescope Science Institute, which is
operated by the Association of Universities for Research in Astronomy, Inc.,
under NASA contract NAS 5-26555. These observations are associated with
programs \#11700 and 11702.}
\altaffiltext{1}{Center for Cosmology and AstroParticle Physics, The Ohio State 
University, 191 West Woodruff Avenue, Columbus, OH 43210, USA}
\altaffiltext{2}{Spitzer Science Center, California Institute of Technology,
MS 220-6, Pasadena, CA 91125, USA}
\altaffiltext{3}{School of Earth and Space Exploration, Arizona State
University, Tempe, AZ 85287, USA}
\altaffiltext{4}{Observatories of the Carnegie Institution of Washington,
813 Santa Barbara Street, Pasadena, CA 91101, USA}
\altaffiltext{5}{Astronomy Department, The University of Arizona, Tucson,
AZ 85721, USA}
\altaffiltext{6}{Leiden Observatory, University of Leiden, P.O. Box 9513,
Leiden 2300 RA, The Netherlands}
\altaffiltext{7}{Space Telescope Science Institute, 3700 San Martin Drive,
Baltimore, MD 21218, USA}
\altaffiltext{8}{Astronomy Department, California Institute of Technology,
MS 249-17, Pasadena, CA 91125, USA}
\altaffiltext{9}{Physics Department, The University of Arizona, Tucson,
AZ 85721, USA}
\altaffiltext{10}{Hubble Fellow}

\begin{abstract}

  We report the first results from the Hubble Infrared Pure Parallel Imaging
Extragalactic Survey, which utilizes the pure parallel orbits of the 
{\it Hubble Space Telescope} to do deep imaging along a large number of random
sightlines. To date, our analysis includes 26 widely separated fields observed
by the Wide Field Camera 3, which amounts to 122.8 arcmin$^2$ in total area.
We have found three bright $Y_{098}$-dropouts, which are candidate galaxies
at $z\gtrsim 7.4$. One of these objects shows an indication of peculiar
variability and its nature is uncertain. The other two objects are among
the brightest candidate galaxies at these redshifts known to date ($L>2L^*$). 
Such very luminous objects could be the progenitors of the high-mass Lyman
break galaxis (LBGs) observed at lower redshifts (up to $z\sim 5$). While our
sample is still limited in size, it is much less subject to the uncertainty
caused by ``cosmic variance" than other samples because it is derived using
fields along many random sightlines. We find that the existence of the
brightest candidate at $z\approx 7.4$ is not well explained by the current
luminosity function (LF) estimates at $z\approx 8$. However, its inferred
surface density could be explained by the prediction from the LFs at
$z\approx 7$ if it belongs to the high-redshift tail of the galaxy population
at $z\approx 7$.

\end{abstract}

\keywords{cosmology: observations --- galaxies: luminosity function, mass function}

\section{Introduction}

  The Wide Field Camera 3 (WFC3) on board the {\it Hubble Space Telescope (HST)}
has begun breaking new grounds in our exploration of the early universe. Using
the ultra-deep observations taken by the ``HUDF09'' program (PI: Illingworth),
samples of candidate galaxies at $z\approx 7$ have been greatly enlarged
(Oesch et al. 2010; Bunker et al. 2010; McLure et al. 2010; Yan et al. 2010a),
and candidates have been found at $z\approx 8$ (Bouwens et al. 2010b; 
McLure et al. 2010; Yan et al. 2010a) and possibly out to $z\approx 10$
(Yan et al. 2010a; Bouwens et al. 2010c). The studies of the stellar 
populations of galaxies at $z\gtrsim 7$ are now underway (e.g., Bouwens et al.
2010a; Labb\'{e} et al. 2010a, 2010b; Finkelstein et al. 2010). The pencil-beam
HUDF09 data are augmented
by the shallower but $8\times$ wider observations (Windhorst et al. 2010) of
the WFC3 Early Release Science Program 2 (ERS2; PI: O'Connell), which provide
us a unique opportunity to study the bright end of the luminosity function
(LF) at $z\gtrsim 7$ because of the much increased sample size of bright
candidates (Wilkins et al. 2010a, 2010b; Bouwens et al. 2010d; 
Yan et al. 2010b).
Even with the ERS2 data, however, we are only barely able to probe the
bright end at $L\sim L^*$. It is important to extend our investigation to the
brighter regime, as it could be where the progenitors of the high stellar-mass 
($\gtrsim 10^{10}M_\odot$) galaxies at $z\approx 6$--5 would locate
(Yan et al. 2006). Such studies will need to survey larger areas, and
preferably along different sightlines in order to reduce the bias caused by the
underlying large scale structures (a.k.a. ``cosmic variance"; see, e.g.,
Robertson 2010). 

   Our Hubble Infrared Pure Parallel Imaging Extragalactic Survey, or
``HIPPIES'', was designed for this purpose. This program utilizes the unique
``pure parallel" observing mode of {\it HST}, which enables simultaneous
observations using instruments that are not doing the primary observations.
As the pointings of the primary observations are drawn from a wide variety of 
programs of different objectives, the associated parallel observations are in 
effect observing completely random fields, and hence are ideally
suited for minimizing the impact of cosmic variance. In {\it HST} Cycle 17,
there were two pure parallel programs using WFC3 as the imaging instrument, one
led by our team (PID 11702, PI: H. Yan) and the other led by M. Trenti et al.
(PID 11700). Our program has been extended to the coming Cycle 18 (PID 12286).
HIPPIES will incorporate the data from all these programs for its science
objectives, one of which being addressing the very bright-end of galaxy LF at
$z\approx 7$ and beyond. Here we report our initial results based on the data
taken by the end of Cycle 17. Throughout this Letter, we use the following
cosmological parameters: $\Omega_M=0.27$, $\Omega_\Lambda=0.73$ and 
$H_0=71$~km~s$^{-1}$~Mpc$^{-1}$. The quoted magnitudes are all in the AB
system.

\section{Data Description}

   In Cycle 17, pure parallels were only allowed when the prime instrument was
either the Cosmic Origins Spectrograph (COS) or the Space Telescope Imaging
Spectrograph (STIS). Programs 11700 and 11702 only requested the high galactic
latitude opportunities at $|b|>20^{\rm o}$, and only
requested long-duration opportunities ($\gtrsim 3$~orbits) so that the
observations could be sufficiently deep. The observations and data from these
two programs are described below.

\subsection{Observations}

   These two programs used
both the infrared (IR) and the UV-optical (UVIS) channels of WFC3. The IR
images were taken in the same three bands in both programs, namely, 
F098M, F125W, and F160W (hereafter $Y_{098}$, $J_{105}$, and
$H_{125}$, respectively). In the UVIS channel, program 11700 used F606W
($V_{606}$), while program 11702 used F600LP ($LP_{600}$). The left panel of
Figure 1 shows the system response curves in these passbands.

   Our current work includes 26 WFC3 pure parallel fields acquired by the
end of Cycle 17 (2010 August). The total exposure time in these fields ranges
from 6 to 34 ks, with the median of 10.8 ks ($\sim$ 5 orbits).
The distribution of exposure time in the four bands roughly follows the ratio of
UVIS:F098M:F125W:F160W$=1:2:1:1$, but varies significantly among the fields.
The CCD observations in UVIS  always have at least two CR-SPLIT exposures that
enable the rejection of cosmic rays. Most of the IR observations also have at
least two exposures each band in each field. As the IR array was always
non-destructively read-out multiple times during the exposure
(at least five times for these observations), cosmic ray rejection generally
is not a problem, even with single exposures.

\subsection{Data Reduction}
  
   Our data reduction starts from the outputs from the {\it HST}
On-the-Fly Reprocessing (OTFR) pipeline, which are photometrically
calibrated images in units of count rate (i.e., the ``*\_flt.fits''
images fetched from the {\it HST} data archive). 

  The vast majority of our parallel observations are non-dithered because the 
primary COS or STIS observations are not dithered. This makes it difficult to
reject image defects using the usual outlier clipping method in the stacking
process. Most of the defects are flagged in the ``data quality'' (DQ) 
extension of the processed image, which records the pixels that are diagnosed
by the processing software to be problematic for various reasons (e.g., bad
pixels, problems in telemetry, unstable response, etc.). In
order to reduce their chance of contaminating our dropout sample, such image
defects in the IR were filled by interpolation of nearby pixels. Using
the archival ERS2 data, we also identified several thousands of additional
outlier pixels that deviate from linear response at 5$\sigma$ level, and these
pixels were also fixed in the same way. The defected pixels were not treated
for the UVIS images in the current work, because these images are only used for
the ``veto'' purpose in the dropout selection, and hence such image defects in
UVIS do not likely cause any contamination.

   The MultiDrizzle software (Koekemoer et al. 2002) was then used to correct
for the geometric distortions and to stack the images in each band. For each
field, the World Coordinate System (WCS) of the first image in $Y_{098}$ was
always chosen as the reference to align all images. The final pixel scale that
we adopted is $0\farcs 09$~pixel$^{-1}$. The pixel noise of the drizzle-combined
images is correlated at small scale because of the subpixel sampling. We
followed the procedure utilized in the GOODS program (Dickinson et al. 2004)
to calculate the correlation amplitudes in our mosaics, and then to derive the
so-called ``RMS maps" that register the absolute root-mean-square noise in each
pixel.

\subsection{Photometry}

  Matched-aperture photometry was carried out by running the SExtractor program
of Bertin \& Arnouts (1996) in dual-image mode. The $J_{125}$-band mosaics were
used as the detection images. We adopt the SExtractor MAG\_AUTO magnitudes, 
which were measured using the default (Kron factor, minimum radius) of
(2.5, 3.5). A 2-pixel, $5\times 5$ Gaussian filter was used to convolve the
$J_{125}$-band images for detection; the detection threshold was set to
1.5$\sigma$, and a minimum of four connected pixels above the threshold was
required. The RMS maps were used to measure the noise in the pixels within the
MAG\_AUTO apertures. We only include sources at S/N$\geq 5$ in the
$J_{125}$-band (within the MAG\_AUTO aperture) for further analysis.

  The depth of our mosaics varies significantly because the exposure time and
the background condition vary significantly among the HIPPIES fields. For
reference, the median 5$\sigma$ depths measured within a $0\farcs 2$-radius
aperture are 27.36, 27.08, 27.16, 27.05, and 26.68 mag in $V_{606}$, $LP_{600}$,
$Y_{098}$, $J_{125}$, and $H_{160}$, respectively. The source count at
$J_{125}\leq 25.2$~mag is complete in every field, and is complete at
$J_{125}\leq 26.0$~mag for 20 out of the 26 fields.

\begin{figure}[tbp]
\centering
\includegraphics[width=\linewidth]{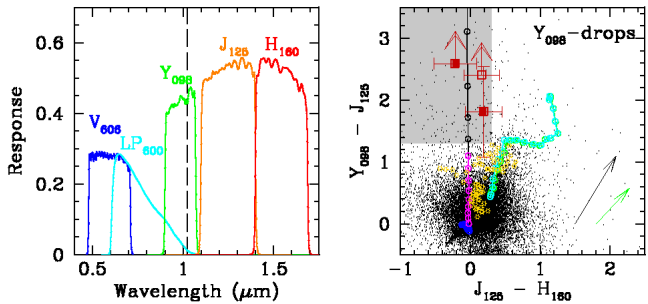} 
\caption{(left) System response curves of the WFC3 passbands used in the current
pure parallel observations. Both programs used the same filters in the IR.
In the UVIS, program 11700 used $V_{606}$ while program 11702 used $LP_{600}$.
The vertical dotted line indicates the location of Ly${\rm \alpha}$ line
redshifted to $z=7.4$.
(right) Color-color diagram demonstrating the selection of $Y_{098}$-dropouts.
The black dots are the field objects. The shaded area indicates the selection
area, where the two objects in our final sample are shown in solid red squares
and the peculiar $Y_{098}$-dropout that has an indication of variability is
shown in open red square. The connected circles to left show the color track of
a typical young galaxy at high redshifts using the model from Bruzual \&
Charlot (2003 (BC03); $\tau=0.3$~Gyr and age of 100~Myr), and attenuated by the
intervening H I absorption along the sightline using the formalism of 
Madau (1995). Different colors are used to code different redshift
ranges along these tracks: blue is for $z<6.4$, magenta is for
$6.4\leq z\leq 7.3$, and black is for $z\geq 7.4$. The selection area is well
separated from the regions occupied by the potential sources of contaminations.
The yellow stars show the colors of Galactic brown dwarfs (e.g. 
Leggett et al. 2002). The green symbols show the colors of a typical red galaxy
at $z\approx 1$--3 simulated using a BC03 model ($\tau=50$~Myr and age of
2.0~Gyr). The cyan symbols represent the colors of the same red galaxy but with
a strong [OIII] emission line of restframe equivalent width of 100\AA. The
black and the green arrows represent the reddening vectors in these passbands
appropriate for $z=7.4$ and $z=2$, respectively, using the reddening law of
Calzetti (2001) and $E(B-V)=0.5$~mag.
}
\end{figure}

\section{Candidate Galaxies at $z\gtrsim 7.4$}

  In this section, we describe the selection of candidate galaxies at
$z\gtrsim 7.4$ as $Y_{098}$-dropouts using HIPPIES data. 
  
\subsection{$Y_{098}$-dropout selection}

    Because $Y_{098}$ does not overlap with $J_{125}$, it is possible to adopt
a large color decrement such that the selection is less prone to the confusion
with the 4000\AA\, break of red galaxies at low redshifts and less affected by
photometric errors. Following Yan et al. (2010b), our main color criteria are
$Y_{098}-J_{125}> 1.3$ mag and non-detections (i.e., S/N$<2$) in the ``veto''
$V_{606}$ and/or $LP_{600}$ images. Based on our simulation using a large
number of galaxy templates from the models of Bruzual \& Charlot (2003) and the
line-of-sight H I absorption recipe of Madau (1995), the adopted
$Y_{098}-J_{125}$ color decrement is appropriate for selecting galaxies at
$z\gtrsim 7.4$. As our UVIS images are not very deep, and
we do not have data at longer wavelength (such as {\it Spitzer} IRAC data;
see, e.g. Yan et al. 2010a, 2010b),
we further require that a legitimate candidate should have 
$J_{125}-H_{160}<0.3$ mag in order to further reduce the chance of 
contamination from red galaxies at low redshifts.
The right panel of Figure 1 shows the selection of $Y_{098}$-dropouts on the
$J_{125}-H_{160}$ versus $Y_{098}-J_{125}$ color-color diagram. 

   The selected candidates were visually inspected to exclude objects that were
formally reported to have S/N$<2$ in the UVIS image but still visible to eyes.
In this step, we also rejected objects that were caused by artifacts and image
defects that are not identified in the data reduction process (see Section 2.2).
A last concern about the contamination is that the interpolation of the
identified defected pixels might leave some residuals that could mimic the
colors of $Y_{098}$-dropouts by chance. To deal with this problem, we created
masks of the interpolated pixels (see Section 2.2) for the individual images,
and ran MultiDrizzle to project them to the same reference frame of the science
images. The locations of the candidates were examined on these projected masks,
and we only kept the objects that do not have any defects in the
central $6\times 6$ pixel area.

\begin{figure}[tbp]
\centering
\includegraphics[width=\linewidth]{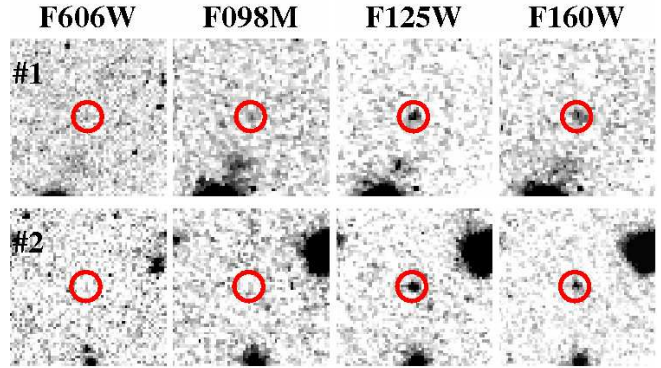} 
\caption{Image stamps of the two $Y_{098}$-dropouts in our sample. 
Images are $5\farcs 4\times 5\farcs 4$ in size. The
red circles are $0\farcs 5$ in radius. North is up and East is to left.
}
\end{figure}

\subsection{Sample}

   Our selection above resulted in a total of three $Y_{098}$-dropouts over 26
HIPPIES fields that amount to 122.8 arcmin$^2$ in area. The colors of these
three $Y_{098}$-dropouts are superposed in the right panel of Figure 1
(solid red squares). Their image are shown in Figures 2 and 3,
and Table 1 lists their photometric information.

\begin{figure}[tbp]
\centering
\includegraphics[width=\linewidth]{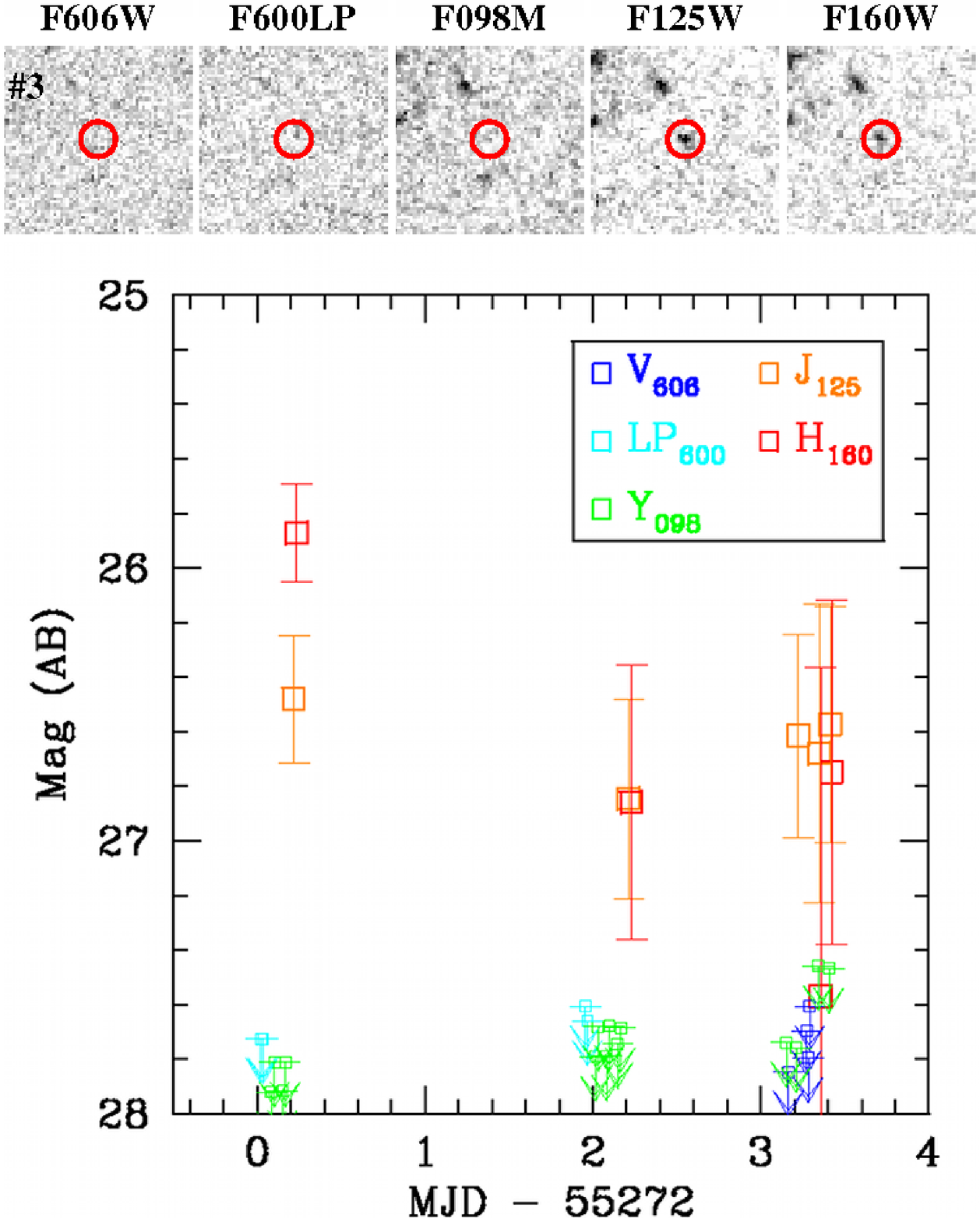}
\caption{(top) Same as Fig. 2, but for the variable object
(YsDrop03). The field was observed by both programs 11700 and 11702, and hence
has UVIS data in both $V_{606}$ and $LP_{600}$.
(bottom) 
Light curves of HIPPIES YsDrop03. The $J_{125}$ and $H_{160}$ data
points are matched-aperture MAG\_AUTO magnitudes obtained using the master
$J_{125}$ stack as the detection image, while the arrows are the 2$\sigma$ 
upper limits measured within an $r=0\farcs2$ aperture. The errors are 
1$\sigma$ error within the MAG\_AUTO aperture.
}
\end{figure}

   One of these three objects, YsDrop03 (see Figure 3), shows an indication of
peculiar variability, which makes its candidacy as a high-redshift galaxy 
less certain.  We performed photometry in each individual image, and obtained
its light curves as shown in Figure 3. The variability is the most obvious in
$H_{160}$, where it is the brightest in the first image, declines by
$\sim 1.0$~mag within 2 days, and remains nearly constant in the following 
$\sim 1.2$ days. Similar trend is also seen in $J_{125}$, albeit less
significantly. However, such a variability does not seem to present in $Y_{098}$
or the two UVIS bands, where this object always remains undetected although 
the observations in these bands were executed both before and after
the first images were taken in $H_{160}$ and $J_{125}$. Therefore, this object
still remains a legitimate $Y_{098}$-dropout after this scrutiny. If it is not
at $z\gtrsim 7.4$, the only viable explanation for its color and its variability
is that it is a flaring brown dwarf star within our Galaxy (e.g., Schmidt et al.
2007). However, this interpretation is less favored, because this object has
full width at half maximum (FWHM) of $\sim 0\farcs 3$ in both $J_{125}$ and
$H_{160}$ stacks and hence seems to be resolved. Another possibility is that
this object is an active galactic nucleus (AGN; see, e.g., Cohen et al. 2006)
at $z\gtrsim 7.4$, or a transient at a similar redshift. 
Without further data, it is difficult to determine its nature. In any case, it
does not fit in the conventional picture of an LBG, and therefore we do not
include it in our final sample.

  The remaining two are good candidate galaxies at $z\gtrsim 7.4$. None of them
show time variability. Both are resolved (FWHM $\gtrsim 0\farcs 3$--$0\farcs 4$)
in $J_{125}$ and $H_{160}$ bands and hence are not likely brown dwarfs. 
Finally, they are not likely red galaxies at $z\approx 1$--3 whose 4000\AA\,
break could mimic Lyman break, as their $J_{125}-H_{160}$ colors are much bluer
than those of such interloppers. This is particularly true for YsDrop02, 
which has a very blue color of $J_{125}-H_{160}\approx -0.2$~mag. A few similar
cases have already been observed in candidate galaxies at $z\approx 7$ and
beyond (e.g., Oesch et al. 2010; Bouwens et al. 2010a; Yan et al. 2010a, 2010b).

\begin{figure}[tbp]
\centering
\includegraphics[width=\linewidth]{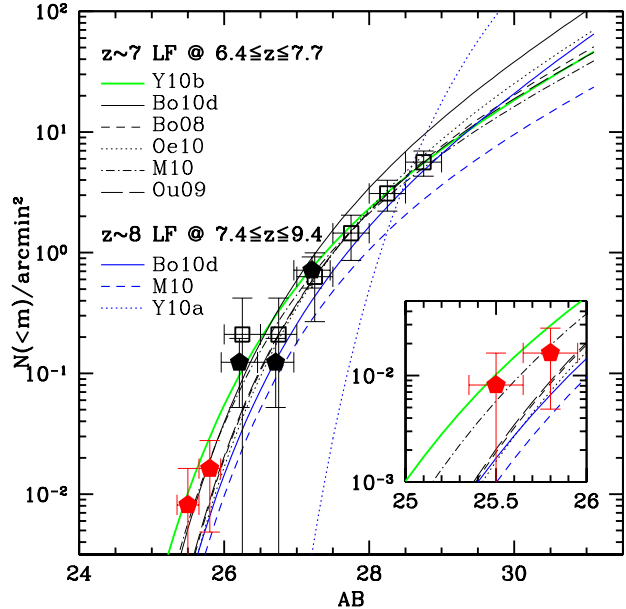} 
\caption{
Comparison of the surface density of $Y_{098}$-dropouts at the very bright
end inferred from our sample (red solid pentagons) and the predictions from
various LFs at $z\approx 7$ and 8 (curves). The inset zooms in the region at 
the very bright end. The $z\approx 7$ LFs are taken from 
Bouwens et al. (2008; Bo08), Ouchi et al. (2009; Ou09), McLure et al.
(2010; M10), Oesch et al. (2010; Oe10), Bouwens et al. (2010d; Bo10d), and 
Yan et al. (2010b; Y10b), while the $z\approx 8$ LFs are taken from Bo10d, M10
and Yan et al. (2010a; Y10a). The $z\approx 7$ LFs are integrated over
$6.4\leq z\leq 7.7$, while the $z\approx 8$ ones are integrated over
$7.4\leq z\leq 9.4$. The surface density of galaxies at $z\approx 7$
inferred from the $z_{850}$-dropouts in the HUDF (black open squares; Yan et al.
2010a) and the ERS2 field (black solid squares; Yan et al. 2010b) are also
plotted.
}
\end{figure}

\section{Discussion}

  The two $Y_{098}$-dropouts in our final sample are the brightest candidate
galaxies at $z>7$ discovered by WFC3 to date. Even assuming that their
redshifts are at the lower end of the selection window ($z\approx 7.4$), we
obtained their absolute magnitudes of $M=-21.6\pm 0.16$ and $-21.2\pm 0.15$~mag
for YsDrop01
and YsDrop02, respectively, which are at least $2\times$ brighter than any
current estimates of the $M^*$ value at $z\approx 7$ or 8. Earlier surveys
over contiguous fields have resulted in a number of $z$-band dropouts
(candidate galaxies at $z\approx 7$) of comparable brightness (e.g.,
Ouchi et al. 2009; Hickey et al. 2010; Castellano et al. 2010) or even brighter
(Capak et al. 2009), and our current work indicates that such high luminosity
galaxies could exist at $z\gtrsim 7.4$.

  Using the formalism of Madau et al. (1998), the rest-frame UV luminosities of
our candidates correspond to star formation rates (SFRs) of $23.7^{+3.5}_{-3.1}$
and $16.4^{+2.4}_{-2.1}$~$M_\odot$~yr$^{-1}$. As an 
order-of-magnitude estimate, if they keep forming stars at the same rates to
$z\approx 5$, they could accumulate stellar masses of 
0.7--$1.1\times 10^{10}M_\odot$ in the next $\sim 473$ million years. Therefore,
our $Y_{098}$-dropouts could be the progenitors of the
high-mass LBGs observed at $z\approx 5$ that have stellar masses to the order
of $\sim 10^{10} M_\odot$ (see e.g., Yan et al. 2005; Stark et al. 2007).

  The existing LF estimates of LBGs at $z\approx 7$ and beyond are largely
based on observations at $L\lesssim L^{*}$. For this reason, it is interesting 
to see whether the number density of such very bright candidate galaxies at 
$z\gtrsim 7.4$ is consistent with the expectations from such LFs. This
comparison is shown in Figure 4, where the result from this study is shown as
the red pentagons.
As the formal redshift selection window of $Y_{098}$-dropout spans 
$7.4\lesssim z\lesssim 9.4$, we compare to the predictions from the LFs at both
$z\approx 8$ and 7. Our data point is apparently higher than the predictions
from the $z\approx 8$ LFs, including the one of Bo10d that
predicts the highest surface density. If we single out the brightest object
YsDrop01, the discrepancy is more obvious. The Bo10d $z\approx 8$ LF predicts a
cumulative surface density of $\sim 2.7\times 10^{-3}$ per arcmin$^{2}$ to
$\leq 25.6$~mag over $7.4\lesssim z\lesssim 9.4$, which is a factor of three
lower than the density inferred from our YsDrop01 ($8.1\times 10^{-3}$ per
arcmin$^{-2}$). On the other hand, the LF of Yan et al. (2010b;
$M^{*}=-20.33$, $\alpha=-1.80$ and $\Phi=5.52\times 10^{-4}$~Mpc$^{-3}$)
predicts the highest bright-end surface density at $z\approx 7$. This LF, like
others that are mostly based on the WFC3 $z_{850}$-dropout results, is
applicable over the redshift range of $6.4\lesssim z\lesssim 7.7$. It predicts
a cumulative surface density of $\sim 1.5\times 10^{-2}$ per arcmin$^{2}$ to
$\leq 25.6$~mag, which seems to be able to explain the existence of YsDrop01
if it is at $z\lesssim 7.7$. 

\section{Summary}

   In this Letter, we present the initial results from the HIPPIES program. To
date, our analysis includes 26 widely separated {\it HST} WFC3 pure parallel
fields at $|b|>20^{\rm o}$ obtained in Cycle-17, which amount to 
122.8~arcmin$^2$ in total area. Using these data, we search for candidate 
galaxies at $z\gtrsim 7.4$ as $Y_{098}$-dropouts. One candidate that we found
shows an indication of peculiar variability in $J_{125}$ and $H_{160}$ but
remains undetected in the bluer bands, and its nature is unclear. Excluding
this object, our current sample consists of two very bright candidates at
$L>2L^{*}$, which, based on their SFR estimates, could be linked to the
progenitors of high-mass LBGs observed at lower redshifts (up to $z\sim 5$).
While its size is still very limited, our
sample is constructed from a large number of random fields and thus the impact
of cosmic variance is minimal. The surface density inferred from our brightest
candidate is not well explained by the existing LF
estimates at $z\approx 8$, but could be explained by the prediction of LFs
at $z\approx 7$ if it belongs to the high-redshift tail of the $z\approx 7$
galaxy population. Our new HIPPIES
data soon to be taken in the coming {\it HST} cycle ($\sim 42$ pointings) will
enable us to search for different $Y_{105}$-dropouts in a similar way, and the
comparison of the two results will shed light to the evolution of the star
formation activities in the early universe.

\acknowledgements

We thank the referee for the helpful comments.
We acknowledge the support of NASA grant HST-GO-11702.*. 
HY is supported by the long-term fellowship program of the Center for
Cosmology and AstroParticle Physics (CCAPP) at The Ohio State University.
BER is supported by Hubble Fellowship Program number HST-HF-51262.01-A.
RAW is supported by NASA JWST Interdisciplinary Scientist grant NAG5-12460
from GSFC. We dedicate this Letter to the memory of John Huchra, who during his
life has been a very staunch supporter of the {\it Hubble Space Telescope}
project.

\begin{deluxetable}{ccccccccc}
\tablewidth{0pt}
\tablecolumns{9}
\tabletypesize{\scriptsize}
\tablecaption{Properties of $Y_{098}$-Dropouts\tablenotemark{a}}
\tablehead{
\colhead{ID} &
\colhead{R.A. and Decl. (J2000)\tablenotemark{b}} &
\colhead{$V_{606}$} &
\colhead{$LP_{600}$} &
\colhead{$Y_{098}$} &
\colhead{$J_{125}$} &
\colhead{$H_{160}$} &
\colhead{Y$-$J} &
\colhead{J$-$H} 
}
\startdata

HIPPIES-YsDrop01 & 16:31:35.24 +37:36:14.03 & $>28.1$ &    ...  &  27.33$\pm$0.71 & 25.52$\pm$0.16 & 25.33$\pm$0.21 &  1.81  &  0.19 \\
HIPPIES-YsDrop02 & 14:36:50.56 +50:43:33.69 & $>28.4$ &    ...  &  $>28.4$        & 25.86$\pm$0.15 & 26.07$\pm$0.27 & $>2.5$ & $-0.21$\\
HIPPIES-YsDrop03\tablenotemark{c} & 07:50:51.41 +29:16:17.54 & $>28.4$ & $>28.3$ &  $>29.0$        & 26.62$\pm$0.17 & 26.46$\pm$0.20 & $>2.4$ &  0.16 \\

\enddata

\tablenotetext{a.}{All magnitude limits are 2$\sigma$ limits measured within 
an $r=0\farcs 2$ aperture.}
\tablenotetext{b.}{The quoted coordinates are based on the astrometric
solutions provided by the OTFR pipeline for the first $Y_{098}$ image in each
field.}
\tablenotetext{c.}{This object shows indication of variability and thus its
nature is unclear.}
\end{deluxetable}

\end{document}